\documentclass[12pt]{article}
\newcommand{\be}{\begin{equation}}
\newcommand{\ee}{\end{equation}}
\newcommand{\bea}{\begin{eqnarray}}
\newcommand{\eea}{\end{eqnarray}}

\begin{document}
\begin{titlepage}

%\flushright{To Appear: }

\vspace{1in}

\begin{center}
\Large
{\bf Running of the Scalar Spectral Index and Observational Signatures 
of Inflation}

\vspace{1in}

\normalsize

\large{James E. Lidsey$^1$ \& Reza Tavakol$^2$}

\normalsize
\vspace{.7in}

{\it Astronomy Unit, School of Mathematical 
Sciences,  \\ 
Queen Mary, University of London, Mile End Road, LONDON, E1 4NS, U.K.}

\end{center}

\vspace{1in}

\baselineskip=24pt
\begin{abstract}
\noindent Some of the 
consequences for 
inflationary cosmology of 
a scale dependence (running) 
in the tilt of the 
scalar perturbation spectrum are considered. 
In the limit where the running is itself approximately 
scale-invariant, 
a relationship is found between the 
scalar and tensor perturbation amplitudes, 
the scalar spectral index and its running. 
This relationship is 
independent of the functional form of the inflaton potential. 
More general settings, including
that of braneworld cosmological models, are also considered. 
It is found that for the Randall-Sundrum single braneworld scenario,  
the corresponding relation between the observables 
takes precisely the
same form as that arising in the standard cosmology. 
Some implications of the observations failing to satisfy such a 
relationship are discussed. 
\end{abstract}

PACS NUMBERS: 98.80.Cq

\vspace{.7in}
Electronic mail: $^1$J.E.Lidsey@qmul.ac.uk, $^2$R.Tavakol@qmul.ac.uk
 
\end{titlepage}

%\double 

\section{Introduction}

\setcounter{equation}{0}

The Wilkinson Microwave Anisotropy 
Probe (WMAP) has measured the power spectrum of the 
Cosmic Microwave Background (CMB) for multipoles up to 
$l \approx 800$ with unprecedented accuracy \cite{bennett}. The best--fit 
model to the WMAP data alone is consistent with a 
spatially flat universe, with near scale--invariant, adiabatic  
and Gaussian distributed primordial density (scalar) perturbations
\cite{spergel,peiris,komatsu}, 
as predicted by the simplest models of inflation \cite{perturbations}. 
(For a review, see, e.g., Ref. \cite{ll}). 

Combinations of CMB results with other astrophysical 
observations have led to strong constraints on 
the standard cosmological parameters such as the Hubble parameter, 
baryon density and age of the universe \cite{spergel}. 
In order to 
differentiate between the numerous inflationary 
models, however, it is necessary to constrain the power spectrum of 
the primordial fluctuations. Assuming that such a spectrum varies in 
a suitable fashion, the standard approach is to expand 
its logarithm  as a Taylor series in $\ln k$, 
about a given scale, $k_0$: 
\begin{equation} 
\label{expansion}
\ln A_S (k) = \ln A_S (k_0) +(n_S -1)\ln \frac{k}{k_0} 
+\frac{1}{2} \alpha_S \ln^2
\frac{k}{k_0} + \ldots  ,
\end{equation}
where $k$ is the comoving wavenumber, $n_S$ is the spectral index (tilt) 
of the spectrum
and the second--order term, $\alpha_S \equiv (dn_S / d \ln k )_{k_0}$, 
represents the `running' of the spectral index \cite{kt}. 
The `power-law' approximation is equivalent 
to truncating the spectrum to first order, i.e., specifying $\alpha_S=0$. 
At this level of approximation, 
the best--fit to the WMAP data is $n_S=0.99\pm 0.04$ \cite{spergel}. 
On the other hand, there is some evidence that 
the power--law approximation may be inadequate when 
data sets spanning a much wider range of 
scales are combined. 
Specifically, Peiris {\it et al.} \cite{peiris} include CMB data from 
the CBI \cite{cbi} and ACBAR \cite{acbar}
(covering the range of multipoles $800 < 
l< 2000$ complementary to WMAP), together with the 
two degree field (2dF) galaxy redshift survey \cite{2df}
and Lyman--$\alpha$ forest 
data at wavenumbers above $k \approx 0.1 \, {\rm Mpc}$ \cite{forest}. 
There is marginal $1.3 \sigma$ support for a non--zero running, 
$\alpha_S = - 0.055^{+0.028}_{-0.029}$ \cite{peiris}. However, the 
validity of employing the Lyman--$\alpha$ forest data has been 
questioned \cite{seljak}. Bridle {\it et al.} \cite{bridle}
include CMB data from the VSA \cite{vsa}
but not from the Lyman--$\alpha$ forest, 
and find the marginalised 1$\sigma$ result 
$\alpha_S=-0.04\pm 0.03$, in agreement  with the WMAP 
collaboration, although they conclude that 
evidence for a non--zero running is dependent 
on the surprisingly low values of the quadrupole and octopole moments
in the CMB power spectrum. In particular, 
$\alpha_S =0$ is consistent 
when the $l<5$ multipoles are excluded \cite{bridle}. 
Other authors who include only CMB and 2dF data also find that
a scale--invariant tilt is consistent with the observations 
\cite{blm,kkmr,ll03}. 

Although there still remain some open questions 
regarding the interpretation of the data, 
the 
recent developments outlined above provide strong motivation 
for considering what one might expect to learn 
about inflationary cosmology if a running in the spectral index 
is detected \cite{kt,cgl,stewart,models,otherruns,flowrun}. 
This is especially true given the anticipated improvement 
in the 
quality of data from future satellite
experiments such as {\it Planck}.
In general inflationary settings
one would expect the running itself to be scale--dependent. 
However, given the current
absence of observational evidence 
for such a variation and, furthermore, that a
varying running may be 
approximated as a piecewise constant over 
a small enough range of scales, 
we consider inflationary models where 
the running of the tilt is scale--independent and 
non--zero. It is found in Section 2 that the amplitude, tilt and running 
of the scalar spectrum are related in a non--trivial fashion to 
the amplitude of the gravitational wave spectrum that is also 
generated during inflation. A similar relationship is found in Section 
3 for a class of braneworld inflationary cosmologies. We conclude 
with a discussion in Section 4. 

\section{Running of the Spectral Index}

In general, the power spectrum of the scalar perturbations 
is closely related to the 
functional form of the inflaton potential\footnote{We 
employ the normalization conventions of Ref. \cite{llkcba} in this paper.}, 
$V(\phi)$: 
\begin{equation}
\label{scalaramp}
A_S^2= \frac{\kappa^6}{75\pi^2} \frac{V^3}{{V'}^2}  ,
\end{equation}
where $\kappa^2 =8\pi m_P^{-2}$, $m_P$ is the Planck mass and  
a prime denotes $d/d\phi$.
The relationship between the inflaton field and 
comoving wavenumber follows from the scalar field 
equations of motion and is given by 
\begin{equation}
\label{kphi}
\frac{d}{d\ln k} = -\frac{V'}{3H^2} \frac{d}{d \phi}
\end{equation}
in the slow--roll limit. 
By defining the `slow--roll' parameters as \cite{lpb}: 
\begin{eqnarray}
\label{epsilon}
\epsilon \equiv \frac{1}{2\kappa^2} \frac{{V'}^2}{V^2} \\
\label{eta}
\eta \equiv  \frac{V''}{\kappa^2 V} \\
\label{xi}
\xi \equiv \frac{V'V'''}{\kappa^4 V^2} ,
\end{eqnarray}
the spectral index and its running may be expressed directly in 
terms of the potential and its derivatives \cite{kt,ll92}:
\begin{eqnarray}
\label{scalartilt}
n_S-1 = -6 \epsilon +2 \eta \\
\label{running}
\alpha_S= 16 \epsilon \eta - 24 \epsilon^2 -2 \xi  .
\end{eqnarray}
Thus, the running of the spectral index depends on the 
third derivative of the potential. 
Eq. (\ref{scalartilt})
is truncated at order $\xi$, such that 
quadratic corrections in $\epsilon$ and $\eta$ are assumed to be 
negligible. This requires that $| \xi | \ll {\rm max} 
\left( \epsilon , | \eta | \right)$ and is 
equivalent to assuming that $|n_S -1 | \ll 1$ and $| \alpha_S |
\approx  
(n_S -1 )^2$ or less. As emphasized in Ref. \cite{stewart}, 
slow--roll predicts the former condition but not 
necessarily the latter. 

The slow--roll parameter (\ref{epsilon}) 
is also related to the spectrum of tensor (gravitational wave) perturbations 
that are generated quantum mechanically during inflation 
\cite{gw}. The relationship 
is expressed through the consistency equation (for a review, see, e.g., Ref. 
\cite{llkcba}): 
\begin{equation}
\label{consistency}
\frac{A_T^2}{A_S^2} = - \frac{1}{2} n_T , \qquad n_T= -2 \epsilon  ,
\end{equation}
where $\{ A_T , n_T \}$ represent the amplitude and spectral index 
of the tensor perturbations respectively. 

In view of the discussion given in the previous Section,
we consider the case where $\alpha_S$ is assumed to be constant. 
Our aim is to derive an expression relating observable parameters 
in the presence 
of a non-zero running.
This requires the integration of Eq. (\ref{running})
with respect to the inflaton 
field. This equation may be viewed as a third--order, 
non--linear differential equation. 
Its first integral therefore relates the inflaton potential to 
its first two derivatives, or equivalently, the two slow--roll 
parameters (\ref{epsilon}) and (\ref{eta}). Consequently, substitution of
Eqs. (\ref{scalartilt}) 
and (\ref{consistency}) into such an expression then results 
in a constraint equation that relates the 
observable parameters $\{ A_S, n_S, \alpha_S, A_T \}$. 

To proceed, we introduce a new variable
\begin{equation}
\label{defy}
y \equiv \frac{V'}{V}
\end{equation}
representing the logarithmic derivative of the potential.
The third--order equation
(\ref{running}) then reduces to the non--linear, second--order equation
\begin{equation}
\label{runningy}
y y'' - y^2 y' = -\frac{\kappa^4 \alpha_S}{2}  .
\end{equation}
Defining $z \equiv y'$, such that $y'' = z dz/dy$, then reduces 
Eq. (\ref{runningy}) to a first--order equation of the form 
\begin{equation}
\label{runningz}
yz \frac{dz}{dy} - y^2z = -\frac{\kappa^4 \alpha_S}{2}  .
\end{equation}
Eq. (\ref{runningz}) may be further simplified 
by defining the variable
\begin{equation}
\label{defu}
u \equiv z- \frac{1}{2} y^2
\end{equation}
and it follows after substitution of Eq. (\ref{defu}) 
into Eq. (\ref{runningz}) that 
\begin{equation}
\label{yuequation}
\frac{dy}{du} =- \frac{2}{\kappa^4 \alpha_S} \left[ uy +
\frac{1}{2} y^3 \right]  ,
\end{equation}
where we now view $y$ and $u$ as the dependent and 
independent variables, respectively. 
Eq. (\ref{yuequation}) may then be rewritten in a separable form by 
introducing the variable
\begin{equation}
\label{defw}
w \equiv y(u) \exp \left[ \frac{u^2}{\kappa^4\alpha_S} \right]
\end{equation}
and substitution of Eq. (\ref{defw}) into Eq. (\ref{yuequation}) implies that 
\begin{equation}
\label{wu}
\frac{1}{w^3}\frac{dw}{du} = - \frac{1}{\kappa^4\alpha_S}   
\exp \left[ - \frac{2u^2}{\kappa^4\alpha_S} \right]  .
\end{equation}
Eq. (\ref{wu}) 
admits the first integral
\begin{equation}
\label{firstintegral}
\frac{1}{y^2} \exp \left[ -\frac{2u^2}{\kappa^4\alpha_S} \right] -
\left( \frac{\pi}{2\kappa^4 \alpha_S} \right)^{1/2} 
{\rm erf} \left[ \sqrt{ \frac{2}{\kappa^4\alpha_S}} u \right]
= c  ,
\end{equation}
where ${\rm erf} (x) \equiv 2 \pi^{-1/2} 
\int^x_0 ds \, \exp [ -s^2 ]$ represents the error function and 
$c$ is an integration constant. The error function is a monotonically 
increasing function, such that 
${\rm erf} (0) =0$ and ${\rm erf} (\infty ) =1$,
and has a first derivative given by $d \left[ {\rm erf}(x) 
\right] /dx= (2/\sqrt{\pi})\exp (-x^2)$. 
In the case where $\alpha_S <0$, Eq. (\ref{firstintegral}) 
may be expressed in terms of the imaginary error function,  
${\rm erfi} (x) = -i {\rm erf} (ix)$.

Finally, the pair $\{ u , y \}$ may be directly related 
to observable parameters. 
Comparison of Eqs. (\ref{epsilon}) and (\ref{defy}) 
implies that $y^2= 2\kappa^2 \epsilon$ and it 
then
follows from definition (\ref{defu})
that 
the variable $u$ is directly related to the scalar spectral index, 
$u=\kappa^2 (n_S-1)/2$. Thus, 
substituting Eq. (\ref{consistency}) into Eq. (\ref{firstintegral}) 
implies that 
\begin{equation}
\label{consistentrunning}
\frac{A^2_S}{A_T^2}\exp \left[ - \frac{(n_S-1)^2}{2 \alpha_S} \right]
- \left( \frac{2\pi}{\alpha_S} \right)^{1/2} {\rm erf} \left[ 
\frac{n_S-1}{\sqrt{2\alpha_S}} \right] = \tilde{c}  ,
\end{equation}
where $\tilde{c}$ is an undetermined constant that is, in principle, 
measurable. 
Eq. (\ref{consistentrunning}) represents an observable signature 
of inflationary models that generate a scale--invariant running 
of the spectral index. 

%-----------------------------------------
\section{Running and Braneworld Cosmology}
%-----------------------------------------

It is also of interest to investigate whether 
the above type of observable signature holds in
other more general inflationary scenarios.
In recent years, considerable interest has focused on the possibility that 
our observable four--dimensional universe may be viewed as a 
domain wall or `brane' that is embedded in a higher--dimensional `bulk' space
\cite{early,hw,RSII}. 
According to these scenarios, the standard model gauge interactions are confined to the 
brane, but gravity may propagate in the bulk \cite{hw}. The 
motion of the brane through the static bulk space is interpreted 
by an observer confined to the brane as 
cosmic expansion or contraction \cite{kraus}. 

In this Section, we consider the 
Randall--Sundrum type II (RSII) braneworld scenario, where a single brane is 
embedded in five--dimensional anti--de Sitter (AdS) 
space \cite{RSII}. In this case, the effective Friedmann equation on the brane 
is derived from the Israel junction conditions that relate the 
extrinsic curvature of the induced metric on the brane to 
the energy--momentum of the matter fields that are confined to the 
brane \cite{israel}. 
The form of the Friedmann equation is modified from that of standard cosmology 
based on Einstein gravity and acquires a quadratic dependence 
on the energy density, $\rho$ \cite{quadratic}: 
\begin{equation}
\label{RSfriedmann}
H^2= \frac{\kappa^2}{3} \rho \left[ 1 + \frac{\rho}{2\lambda} \right]  ,
\end{equation}
where $\lambda$ is the brane tension. 

Such a modification becomes important 
at high energies and has significant implications 
for inflation \cite{maartens,steep}. In particular, in the limit where 
$\rho \gg \lambda$, the amplitudes of the scalar and tensor 
perturbations are enhanced \cite{maartens,lmw}: 
\begin{equation}
\label{RSspectrum}
A_S^2 = \frac{\kappa^6}{600\pi^2} \frac{V^6}{\lambda^3 {V'}^2} , \qquad 
A_T^2 = \frac{\kappa^4}{200\pi^2}\frac{V^3}{\lambda^2}  .
\end{equation}
However, despite these corrections, 
the consistency equation relating the two spectra 
is {\em identical} to 
that of the standard scenario, Eq. (\ref{consistency}) \cite{gl}. 
Such a degeneracy in the consistency equation also arises in 
more general brane cosmologies \cite{moregeneral}. 

A natural question to address, therefore, is whether such a degeneracy 
may be lifted by allowing for a running of the spectral index. 
In view of the modifications to the Friedmann equation that typically arise 
in brane cosmology, 
we consider 
a model described by a Friedmann equation of the form
\begin{equation}
\label{qfriedmann}
H^2= \frac{\tilde{\kappa}^2}{3} \rho^q  ,
\end{equation}
where $\rho$ is the energy density of the matter, $q$ is an arbitrary, 
positive constant and $\tilde{\kappa}^2$ is an arbitrary 
constant. Eq. (\ref{qfriedmann}) may be viewed as a limiting case of 
a more generalized Friedmann equation that is relevant 
in the high energy regime of early universe dynamics. For example, 
the case $q=2$ corresponds to 
the RSII scenario when the energy density dominates the brane tension  
and we specify $\tilde{\kappa}^2 = \kappa^2/2\lambda$. 
A further case of interest is given by $q=2/3$. A Friedmann equation 
of this form arises in the extended version of the RSII scenario when 
a Gauss--Bonnet combination of curvature invariants is included 
in the five--dimensional bulk action \cite{gb}. 

We further assume that the universe is dominated by a 
single, self--interacting inflaton field. Conservation of energy--momentum 
of this field then implies that 
\begin{equation}
\label{qscalar}
\ddot{\phi} +3H\dot{\phi} +V' =0  ,
\end{equation}
where a dot denotes differentiation with respect to time. 
We define the generalized slow--roll parameters as 
$\epsilon_g \equiv -\dot{H}/{H^2}$, $\eta_g \equiv V''/(3H^2)$ and 
$\xi_g \equiv V'V'''/(3H^2)^2$, respectively, and in 
the slow--roll limit, $\dot{\phi}^2 \ll V$ and $|\ddot{\phi} |
\ll H|\dot{\phi}|$, these 
reduce to 
\begin{eqnarray}
\label{qepsilon}
\epsilon_g = \frac{q}{2\tilde{\kappa}^2} \frac{{V'}^2}{V^{q+1}} \\
\label{qeta}
\eta_g = \frac{V''}{\tilde{\kappa}^2 V^q} \\
\label{qxi}
\xi_g = \frac{V'V'''}{\tilde{\kappa}^4 V^{2q}}  .
\end{eqnarray}

Conservation of energy--momentum implies that the curvature perturbation 
on uniform density hypersurfaces is conserved on super--Hubble radius
scales. 
This follows as a direct consequence of energy--momentum conservation 
of the inflaton and is independent of the gravitational 
physics \cite{wands}. 
It can then be shown that 
the amplitude of the scalar perturbation spectrum is given by 
$A_S^2 \propto H^4/\dot{\phi}^2$ in the slow--roll limit \cite{wands}. 
The value of the scalar field is related to 
the comoving wavenumber through Eq. (\ref{kphi}). 
Substituting the field equations into the expression for the amplitude and 
differentiating with respect to comoving wavenumber then implies 
that the scalar spectral index is given by 
\begin{equation}
\label{qscalartilt}
n_S-1 = -6 \epsilon_g +2 \eta_g
\end{equation}
and the running of the tilt is given by 
\begin{equation}
\label{qrunning}
\alpha_S= 16 \epsilon_g \eta_g -\frac{12(q+1)}{q} \epsilon_g^2 -2 \xi_g  .
\end{equation}

As in the previous Section, our aim is to integrate 
Eq. (\ref{qrunning}) under the assumption that the running of 
the spectral index is constant. 
It proves convenient to define a new scalar field, $\varphi$: 
\begin{equation}
\label{defvarphi}
\frac{d}{d \varphi} \equiv V^{(1-q)/2} \frac{d}{d \phi}
\end{equation}
and this implies that Eq. (\ref{qrunning}) takes the form 
\begin{equation}
\label{runningstar}
\tilde{\kappa}^4 \alpha_S = 4(q+1) \frac{(V^*)^2 V^{**}}{V^3} 
-2(2q+1) \frac{(V^*)^4}{V^4} -2 \frac{{V^*}{V^{***}}}{V^2}  ,
\end{equation}
where a star denotes $d/d\varphi$. Defining the new variable
$Y \equiv V^*/V$ then simplifies Eq. (\ref{runningstar}) to 
\begin{equation}
\label{Yrunning}
Y Y^{**} +(1-2q) Y^2 Y^* = - \frac{\tilde{\kappa}^4\alpha_S}{2}  .
\end{equation}

Eq. (\ref{Yrunning}) can be integrated in a similar way to 
that employed in Section 2 and we therefore omit the details. 
Eq. (\ref{Yrunning}) reduces to the separable equation
\begin{equation}
\label{WU}
\frac{1}{W^3}\frac{dW}{dU} = \left( \frac{1-2q}{\tilde{\kappa}^4\alpha_S} 
\right) 
\exp \left[ - \frac{2U^2}{\tilde{\kappa}^4\alpha_S} \right]  ,
\end{equation}
where 
\begin{eqnarray}
\label{Udef}
U \equiv Y^* +\left( \frac{1-2q}{2} \right) Y^2 \\
\label{Wdef}
W(U) \equiv Y \exp \left[ \frac{U^2}{\tilde{\kappa}^4 \alpha_S} \right]
\end{eqnarray}
and solving Eq. (\ref{WU}) then implies that 
\begin{equation}
\label{qfirstintegral}
\frac{1}{Y^2} \exp \left[ -\frac{2U^2}{\tilde{\kappa}^4\alpha_S} \right] +
\sqrt{\frac{\pi}{2\tilde{\kappa}^4 \alpha_S}} 
(1-2q) {\rm erf} \left[ \sqrt{ \frac{2}{\tilde{\kappa}^4\alpha_S}} U \right]
= C  ,
\end{equation}
where $C$ is an integration constant. 

Comparison of Eqs. (\ref{qscalartilt}) and (\ref{Udef}) implies that 
$U= \tilde{\kappa}^2 (n_S-1)/2$. Moreover, 
substituting Eq. (\ref{defvarphi}) 
into the definition of the variable $Y$ implies that 
$Y^2= (2\tilde{\kappa}^2 /q) \epsilon_g$. It follows, therefore, that Eq. 
(\ref{qfirstintegral}) may be expressed in the form 
\begin{equation}
\label{qfirstint1}
\frac{1}{\epsilon_g} \exp \left[ - \frac{(n_S-1)^2}{2\alpha_S} \right]
+ \left( \frac{2\pi}{\alpha_S} \right)^{1/2} \left( 
\frac{1-2q}{q} \right) {\rm erf} \left[ \frac{n_S -1}{\sqrt{2\alpha_S}} 
\right] = \tilde{C}  ,
\end{equation}
where $\tilde{C}$ is a dimensionless constant. 

In braneworld inflationary scenarios, the calculation of the tensor 
perturbation spectrum is  more involved than that of the 
scalar perturbations because the gravitational waves extend into the
bulk dimensions \cite{lmw}. 
Consequently, one must consider the tensor perturbations 
for each specific model. For the RSII scenario, where $q=2$, 
substituting Eq. (\ref{qepsilon}) into Eq. (\ref{RSspectrum}) 
implies that $A_T^2/A_S^2= 3\epsilon_g/ 2$. Remarkably, 
therefore,  we conclude that 
when Eq. (\ref{qfirstint1}) is expressed in terms of the observables 
$\{ A_S , n_S, \alpha_S, A_T \}$, it reduces to  {\em precisely} the 
same form as the corresponding relationship for the 
standard inflationary cosmology, Eq. (\ref{consistentrunning}).

\section{Discussion}

The inflationary scenario has received a great deal of observational support 
from recent CMB satellite observations \cite{spergel,peiris}. 
From the theoretical 
perspective, an important problem to address is the origin of the inflaton 
field within a fundamental underlying theory and, more specifically, 
the nature of the inflaton potential that drove the 
accelerated expansion of the very early universe. In the 
case where the running of the spectral index vanishes, it is well 
known that the form of the potential leading to a constant 
spectral index is not unique \cite{llkcba,lt}. Indeed, the  
origin of this degeneracy may be understood from a mathematical
point of view by expressing 
the potential in terms of the derivative $V \equiv dW(\phi)/d\phi$
and rewriting
Eq. (\ref{scalartilt}) in the form
\begin{equation}
\label{sch}
\frac{W'''}{W'}-\frac{3}{2} \left( 
\frac{W''}{W'} \right)^2 =\frac{\kappa^2}{2} \left( 
n_S -1 \right)  .
\end{equation}
This novel way of expressing the 
constraint on the potential is illuminating because 
the left hand side of Eq. (\ref{sch}) is the Schwarzian derivative of the 
function $W (\phi)$ \cite{hille}. 
This is the unique elementary 
function of the derivatives that is invariant under the homographic 
transformation that corresponds to the group of fractional linear
transformations:
\begin{equation}
\label{SL2R}
\tilde{W} = \frac{aW +b}{cW+d}  ,
\end{equation}
where $\{ a,b,c,d \}$ are arbitrary constants satisfying 
$ad-bc =1$. Thus, given a particular solution 
to Eq. (\ref{sch}) (such as an exponential potential), 
more general solutions and corresponding potentials 
may be generated by applying the 
transformation (\ref{SL2R}). Further observational input, most notably 
from the gravitational wave background, is required to lift the 
degeneracy \cite{llkcba}. 

In this paper we have considered the more general 
class of inflationary models where 
the running of the scalar spectral index 
is itself  non--zero and independent of scale. In general, it 
is not possible to determine the analytic form of the potentials 
that generate such a spectrum. On the other hand, 
their asymptotic limit may be deduced by noting that 
the first integral of the second--order equation 
(\ref{runningy}) may be written in the form 
\begin{equation}
\label{yf}
y' = \sqrt{f(\phi) -\kappa^4 \alpha_S \ln y}    ,
\end{equation}
where the function $f(\phi)$ itself satisfies the non--linear 
equation, $f' = 2y{y'}^2$, i.e., 
\begin{equation}
\label{fequation}
f' =2y \left( f- \kappa^4 \alpha_S \ln y \right)  .
\end{equation}
The pair of equations (\ref{yf}) and (\ref{fequation}) 
represent a plane autonomous system with a single equilibrium point
located at  
$f = \kappa^4 \alpha_S \ln y$ and it follows 
from the definition of the function 
$f$ that this point represents the asymptotic 
form of the general solution to Eq. (\ref{runningy}) in the limit where
$|y''| \ll y|y'|$, i.e., where the first term on the left hand side of 
Eq. (\ref{runningy}) is negligible. In this limit, 
Eq. (\ref{runningy}) may be integrated to yield the form of the 
potential: 
\begin{equation}
\label{potential}
V=V_0 \exp \left[ \pm \left( 
\frac{81 \left| \alpha_S \right|}{128} \right)^{1/3} \left( \kappa
\phi \right)^{4/3} \right]  ,
\end{equation}
where $V_0$
is an arbitrary positive constant and the 
sign of the exponent corresponds to the sign of the running. 
It is worth remarking that potentials of this specific asymptotic form 
also arise within the context of supergravity models \cite{cnr}.  

Eq. (\ref{consistentrunning})
directly relates the four observable 
parameters $\{ A_S, n_S, \alpha_S ,A_T \}$. 
An important feature of this relation is 
that it is {\it independent} of the specific functional 
form of the inflaton potential. In this sense, therefore, it 
represents an observable signature of the class of inflationary models 
where the `running of the running',  
$\beta_S \equiv 
d \alpha_S /d \ln k$, is negligible. 
However, the expressions for the tilt and running, Eqs. 
(\ref{scalartilt}) and (\ref{running}), were derived under the 
standard assumption that $| \xi | \ll {\rm max} (\epsilon , | \eta |)$.
Although this is consistent, it is not 
required by the slow--roll approximation \cite{stewart} and
failure to satisfy Eq. (\ref{consistentrunning}) may therefore indicate 
that such an assumption would need to be relaxed. 

The occurrence of the arbitrary constant implies that to
proceed observationally the parameters 
$\{ A_S, n_S, \alpha_S ,A_T \}$
must be measured over at least two separate scales. 
One measurement is required to 
determine the numerical value of the integration 
constant and the second to 
determine whether Eq. (\ref{consistentrunning}) is indeed 
satisfied. 
The advantage of Eq. (\ref{consistentrunning}) 
over the consistency equation (\ref{consistency}) is that 
it relates the scalar and tensor perturbation amplitudes directly 
to the scalar spectral index and its running. 
Consequently, it does not require the tilt of the tensor spectrum 
to be directly 
measured. 

On the other hand, we have assumed that the running is 
effectively constant and this 
can only be verified observationally to within some error. 
When discussing observational constraints, the primordial 
power spectrum may be viewed as an unknown function and 
the fitting procedure effectively truncates the Taylor 
expansion (\ref{expansion}) at a finite order. 
This is equivalent to setting all corresponding 
higher--order, slow--roll parameters 
to zero. Establishing whether  
a constant running is a 
good fit to the
data requires the introduction of the next--order term, 
$\beta_S \equiv 
d \alpha_S /d \ln k$, as an additional parameter in the analysis. 
Self--consistency of the assumptions made above 
requires that $\beta_S \approx |n_S-1|^3$ or smaller. 
The assumption that the running is constant would then be 
supported if it turned out that $\beta_S$ has only a 
moderate influence on the likelihood distributions for 
the other observable parameters and is itself entirely consistent with zero
within the observed errors. 

Nevertheless, 
Eq. (\ref{consistentrunning}) may prove important even 
if 
a high running of the running is reported. 
Leach and Liddle have argued that appropriate conditions should 
be satisfied if the inclusion of a
higher--order parameter of the power spectrum is to be justified \cite{ll03}. 
In effect, the criterion is that of convergence in the Taylor expansion
(\ref{expansion}). 
In the present 
context, the inclusion of the running of the running 
could only be justified if the third--order term in 
Eq. (\ref{expansion}) is significantly 
smaller than the second--order term and, quantitatively, this 
requires 
\begin{equation}
\label{smallbeta}
\left| \frac{\beta_S}{3} \ln \left( 
\frac{k}{k_0} \right) \right| \ll \left| \alpha_S \right|  .
\end{equation}
If condition ({\ref{smallbeta}) 
is violated when the 
detection of $\beta_S$ has only a low significance, 
it could be argued that the determination 
of this parameter may be unreliable and that  
it is therefore not appropriate to include it in the analysis \cite{ll03}.  
Indeed, Eq. (\ref{consistentrunning}) proves important in this case because 
it may be employed to yield crucial information about the 
magnitude of the second derivative of the tilt.
Failure of 
Eq. (\ref{consistentrunning}) to satisfy the data (under the assumption 
of a constant running) could be interpreted as evidence 
that the second derivative of the tilt is important 
without the need for allowing this 
parameter to be {\em a priori} non--zero.

We have also considered inflationary models 
that generate a scale--invariant running of the spectral index 
in the Randall--Sundrum type II braneworld scenario. 
Surprisingly, we found that in the high energy limit,  
the relation
(\ref{qfirstint1}), when expressed in terms of the observables
$\{ A_S, n_S, \alpha_S ,A_T \}$,
takes {\em precisely} the
same form as the corresponding relationship for the
standard inflationary cosmology, Eq. (\ref{consistentrunning}).
This provides further evidence
of the degeneracy that exists between the primordial 
perturbations that are generated in the two scenarios even though 
the gravitational physics is manifestly different in the two cases
\cite{gl,moregeneral}. 
It should be borne in mind, however, that 
in these calculations the effects of the bulk space on the evolution 
of the density perturbations has been neglected. This is consistent 
at linear order when considering scalar perturbations of a homogeneous 
background \cite{maartens}. More generally, the backreaction 
perturbs the bulk space away from conformal invariance and generates
a non--trivial Weyl curvature in the bulk \cite{SMS,maartensrev}. 
This plays the role of a non--local energy--momentum source when 
projected down to four--dimensions and thus alters the background dynamics
\cite{SMS,maartensrev,cm}. 
The failure of the relation (\ref{consistentrunning}) 
to be satisfied
in this case could therefore indicate the possible importance of these 
bulk effects.

\vspace{.3in}
\centerline{\bf Acknowledgments}
\vspace{.3in}

JEL is supported by the Royal Society. We thank N. Nunes for 
helpful discussions.

\end{document}